\documentclass{optica-article}
\journal{opticajournal} 
\articletype{Research Article}

\usepackage{lineno}

\usepackage{gensymb}
\usepackage{float}
\usepackage{subcaption}
\usepackage{adjustbox}
\usepackage{chemformula}
\usepackage[capitalise]{cleveref}
\usepackage{booktabs}

\begin{document}

\title{Demonstration of gas thermometry by single-shot coherent Rayleigh--Brillouin scattering in an iron powder flame}

\author{Giliam G.W. van der Wielen\authormark{1,2,$\dag$,*}, Yingjie Zhao\authormark{3,$\dag$}, Conrad J.M. Hessels\authormark{1,2}, Nico J. Dam\authormark{1,2}, and Alexandros Gerakis\authormark{3}}

\address{\authormark{1}Power \& Flow group, Department of Mechanical Engineering, Eindhoven University of Technology, P.O. Box 513, 5600MB Eindhoven, The Netherlands\\
\authormark{2}Eindhoven Institute for Renewable Energy Systems (EIRES), P.O. Box 513, 5600MB Eindhoven, The Netherlands\\
\authormark{3}Scientific Instrumentation and Process Technology, Luxembourg Institute of Science and Technology (LIST), Maison de l’Innovation, 5 Avenue des Hauts-Fourneaux, L-4362 Esch-sur-Alzette, Luxembourg \\
\authormark{$\dag$}The authors contributed equally to this work.\\}

\email{\authormark{*}g.g.w.v.d.wielen@tue.nl} 


\begin{abstract*}
Single-shot coherent Rayleigh--Brillouin scattering (CRBS) with a chirped optical lattice has been applied to a laboratory iron powder flame at atmospheric pressure. From one expression for the CRBS signal, $S \propto \alpha^4(n_0/T)^2I_1I_2I_\mathrm{pr}|G|^2$, four temperature extraction methods are derived and applied to each of the obtained single-shot spectra. These are identified as the signal at zero lattice velocity, the velocity-integrated signal, and fits of the coherent collisionless and full S6 kinetic lineshapes. The two signal-strength methods agree within 2\,\% for every spectrum, while the collisionless fit exceeds the kinetic one by a nearly constant factor of 1.2, somewhat below the bias expected for the collisionless fit under these conditions. The signal at zero lattice velocity requires neither a lineshape model nor transport coefficients, nor, in principle, a spectrally resolved measurement, but it relies on the absolute signal and is therefore sensitive to beam steering and to differences in pulse energy between the reference and flame measurements. The estimated gas temperature near the flame ranges from about 400 to 800\,K and changes by up to several hundred kelvin between consecutive shots, demonstrating the importance of single-shot acquisition for such flames.
\end{abstract*}

\section{Introduction}
\label{sec:introduction}
Iron powder is a promising recyclable, carbon-free energy carrier~\cite{Bergthorson2015DirectCombustion,Bergthorson2018RecyclableMetal}, whose ignition largely depends on the temperature of the surrounding gas~\cite{Abdallah2024IronParticles}, while the formation of nanoparticles and nitrogen oxides increases with the particle and gas temperatures~\cite{Ning2022CriticalTemperature,Prasidha2024EfficientMetal}. To characterise and better understand these processes, accurate gas-temperature measurements are crucial. Thermocouples and other physical probes, however, perturb the flow, suffer from radiation losses and particle deposition, and are too slow to follow an unsteady flame. Laser-based techniques avoid these problems and have been used to measure gas temperatures and species in metal combustion, mostly of aluminium, and in other particle-laden flames: planar laser-induced fluorescence (PLIF) of \ch{AlO} has been implemented to map species and temperature around single, isolated aluminium particles~\cite{Bucher1996FlamesStructure,Bucher1998PLIFSpecies}. PLIF of \ch{Al} atoms has imaged the particles in aluminised solid-propellant flames~\cite{Vilmart2019ImagingAluminum}, while multi-line NO-LIF, with \ch{NO} seeded to the gas, has provided gas-temperature fields in an iron-oxide nanoparticle synthesis reactor~\cite{Feroughi2015LaserbasedSitu}. Laser absorption spectroscopy has quantified \ch{Al}, \ch{AlO}, and \ch{AlH} around single burning aluminium droplets~\cite{Ruan2025QuantitativeMeasurement,Wang2025SpatiotemporalQuantification,Wu2026TransientAlH} and the aluminium vapour in flat aluminium suspension flames~\cite{Soo2017EmissionLaser}. Absorption thermometry has targeted \ch{AlO} in aluminium powder flames~\cite{hinnant2026aluminum}, and \ch{CO} in combustion gases~\cite{tancin2021ultrafast} and in aluminised composite propellants~\cite{ruesch2021characterization}. In dense dust flames, however, laser light--particle interaction may perturb the flame, while extinction, multiple scattering, and luminosity accumulate along the beam and detection paths, rendering quantitative measurements considerably more challenging than around single particles~\cite{Soo2017EmissionLaser,peng2023combustion,yu2024dual}. Moreover, absorption techniques measure a line-of-sight average instead of a spatially resolved field. Above all, both LIF and absorption rely on specific transitions of specific species in the gas phase. Aluminium burns largely in the vapour phase and forms such species, but iron burns predominantly heterogeneously, at the particle surface~\cite{a848b102-6355-3736-9807-5d17f9df2293, HASSE2026106305}. Gas-phase \ch{Fe} and \ch{FeO} are also formed during iron combustion~\cite{Li:21}, but in small amounts compared with the heterogeneous reaction, which limits species-selective laser diagnostics~\cite{LI2026136303}; suitable absorbers or fluorescing species are then only present when tracers are added~\cite{Feroughi2015LaserbasedSitu}. Optical thermometry in iron dust flames has therefore mostly targeted the particle temperature, by hyperspectral pyrometry of single burning particles~\cite{Hameete2024ParticleresolvedHyperspectral} and multispectral pyrometry in iron dust flames~\cite{Hagen2026SituXray}, whereas the gas temperature remains largely uncharacterised.

Techniques that do not rely on a species-specific resonance can avoid this restriction. Holographic interferometry has provided gas-temperature fields of laminar aluminium dust flames, however, as a line-of-sight method, it requires an axisymmetric flame~\cite{Vovchuk1994TemperatureField}. Filtered Rayleigh scattering (FRS) has been applied to two-dimensional gas-temperature measurements in a nanoparticle-forming flame, with the local composition determined by Raman scattering; the FRS images were accumulated over many laser pulses~\cite{muller2014two}. Although the vapour absorption filter suppresses Mie scattering, the dense particle loading, weak spontaneous signal, and strong flame luminosity make a sufficient signal-to-noise ratio (SNR) difficult to reach~\cite{VanDerWielen2026FundamentalsFiltered}. Coherent techniques open a way to overcome these issues. Hybrid femtosecond/picosecond coherent anti-Stokes Raman scattering (fs/ps CARS) has been used to measure the rotational temperature in aluminised propellant flames~\cite{kearney2016temperature}. The rotational Raman lines of all molecules lie within the bandwidth of a single broadband pulse, so pure-rotational CARS measures \ch{N2} and \ch{O2} simultaneously, which are abundant in air-fed flames, without tuning to a particular species. Coherent Rayleigh--Brillouin scattering (CRBS) instead probes the translational motion of all scatterers in the gas, including atoms, which exhibit no rotational Raman spectrum. It does not rely on a species-specific resonance or equilibrium between translational and internal degrees of freedom, and can be applied to any gas-phase sample as long as optical access exists.

In CRBS, two pump beams form an optical lattice that drives a density modulation in the gas, which is read out by a probe beam~\cite{Pan2002CoherentRayleighBrillouin,Pan2003CoherentRayleighBrillouin,Pan2004CoherentRayleighBrillouin}. The signal as a function of the lattice velocity forms the CRBS spectrum, and CRBS thermometry has been demonstrated~\cite{Gerakis2016RemotesensingGas,Bak_2023,Randolph_2026}. A complete CRBS spectrum can be recorded in a single shot in two ways. With broadband pump pulses, all lattice velocities are driven at once, and the signal must be dispersed by a high-resolution element such as a Fabry--Pérot etalon or a virtually imaged phased array~\cite{Senior2026SingleshotPicosecond}. With a frequency-chirped optical lattice, the spectrum is instead recorded in time~\cite{gerakis2013single}, on a single fast photodiode, so that flame luminosity is collected only during the pulse of a few hundred nanoseconds and the background is measured directly before the pulse in the same trace. The coherent signal leaves the probe volume as a directional beam, which allows efficient spatial filtering of flame luminosity and diffuse background scattering. This is particularly important for large-scale metal burners, where spontaneous optical diagnostics suffer from poor signal quality~\cite{peng2023combustion}. Direct interaction between the laser light and the particles, however, remains a limitation for laser diagnostics in particle-laden flows. With single-shot acquisition, shots affected by strong laser light--particle interaction can be identified and discarded, while every remaining shot yields its own spectrum, so that shot-to-shot fluctuations in an unsteady flame are captured rather than averaged out. 

In previous CRBS thermometry studies, the temperature is obtained from the spectral \textit{shape} by fitting a kinetic model~\cite{Pan2004CoherentRayleighBrillouin,Tenti1974KineticModel,Bookey2006NarrowbandCoherent,Senior2026SingleshotPicosecond}. The signal \textit{strength} carries temperature information as well. Gerakis et al.\ showed that the spectrally integrated signal scales with the square of the gas density~\cite{Gerakis2016RemotesensingGas}, which is the coherent counterpart of spontaneous Rayleigh thermometry, where instead the signal scales with the density itself~\cite{Miles2001LaserRayleigh}. In this work, a portable, frequency-agile laser system~\cite{zhao2026compactdualcolourchirpedpulselaser} is used for single-shot CRBS on a laboratory-scale iron powder flame. To the best of our knowledge, this is the first demonstration of single-shot CRBS for translational gas temperature measurements in metal combustion. We show that a single expression for the CRBS signal leads to four methods to extract the temperature from a spectrally resolved spectrum: two that compare the signal strength with a reference measurement at known pressure, and two that fit the lineshape. One of them, based on the signal where the lattice is at rest relative to the gas, i.e.\ at zero lattice velocity, needs neither a lineshape model nor transport coefficients. 

The paper is structured as follows: Section 2 describes the iron burner, the CRBS principle, and the laser system. Section 3 describes the data acquisition and processing, and Section 4 the signal model and the four temperature extraction methods. Section 5 presents and discusses the results, and Section 6 concludes. Appendix A derives the expression for the CRBS signal used in Section 4.

\section{Experimental and CRBS set-up}
\subsection{Metal dust burner}
\begin{figure}[htbp]
    \centering
    \includegraphics[width=0.6\linewidth]{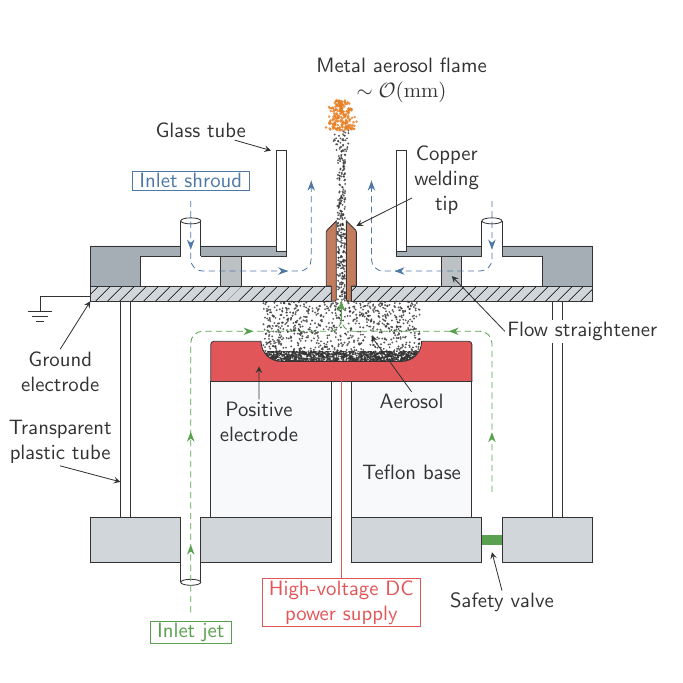}
    \caption{Schematic of the metal dust burner, showing the main components and gas flow paths responsible for particle dispersion and combustion. The metal aerosol flame above the nozzle typically extends over a region of a few millimetres.}
    \label{fig:burnerschematic}
\end{figure}

The experiments used a metal dust burner based on a modified electrodynamic aerosol generator. The design traces its origins to the apparatus of Shoshin and Dreizin~\cite{Shoshin2002ProductionWellControlled}, initially developed to study lifted aerosol flames, and was later adapted by Mohan et al.~\cite{Mohan2005IgnitionAerosolized} to investigate the ignition of single aluminium particles. The present version is closest to the configuration of Ning et al.~\cite{Ning2021BurnTime} but, instead of characterising laser-ignited single iron particles, the goal here is to sustain a continuous iron flame. A schematic of the burner is presented in \cref{fig:burnerschematic}. 

The burner operates on the principle of electrodynamic powder fluidisation. Iron powder is loaded onto the concave bottom electrode of a parallel-plate capacitor enclosed in a chamber. When a high DC voltage is applied, the conductive particles are charged and begin shuttling between the electrodes once the Coulomb force exceeds gravity. Particles are repelled from the positive electrode, collide with the ground electrode, lose their charge, and fall back down. This continuous motion sustains a quasi-steady aerosol cloud between the electrodes. A central gas stream then entrains the aerosolised particles and drives them through the exit nozzle~\cite{Shoshin2002ProductionWellControlled,Ning2021BurnTime}. 

To adapt this system for continuous combustion, several modifications were made. Most notably, commercially available metal inert gas welding tips are used as interchangeable nozzles. Each tip can be replaced to vary the jet diameter, or removed to clear blockages. The thickness of the ground electrode was increased to accommodate the threaded section of the tip, ensuring a secure fit and fixed jet exit position. The capillary-style outlet also acts as a safety buffer, separating the aerosolisation zone from the combustion zone to prevent flashback. 

The shroud flow is guided parallel to the central jet by a short glass tube. Its length was minimised to keep it outside the optical beam paths while maintaining sufficient positional flame stability. Nevertheless, the lifted flame still moves both vertically and radially during operation.

\subsection{Operating conditions} \label{ssec:operating conditions}
The central jet had a diameter of 1\,mm and an exit velocity of 80\,cm/s. The coaxial shroud flow had a diameter of 26\,mm and a velocity of 10\,cm/s. The jet consisted of pure nitrogen, whereas the shroud used a 50/50 nitrogen-oxygen mixture. The dispersed iron powder (EMSURE 1.03819, nominal particle size 10\,$\mu$m) was ignited above the glass tube with a handheld butane lighter. This produced a roughly spherical lifted flame of about 5\,mm diameter that could be maintained for up to 3\,min. All measurements were performed at atmospheric pressure.

\subsection{Principle of single-shot CRBS}
\label{ssec:principle crbs}
CRBS is a seedless four-wave mixing technique that probes the density response of a gas to a moving optical lattice. The spectral shape reflects the molecular velocity distribution and, at high pressures, collective acoustic motion, and yields the temperature~\cite{Pan2002CoherentRayleighBrillouin,grinstead2000coherent}. The signal strength scales with the square of the number density, which gives access to temperature and pressure, as well as the effective polarisability of a gas mixture~\cite{Gerakis2016RemotesensingGas}. A spectral shift arises from a bulk gas flow along the grating vector and gives the flow velocity~\cite{Gerakis2025Velocimetry, Kumar:25}. Unlike spontaneous Rayleigh--Brillouin scattering, CRBS produces a coherent, directional signal beam instead of light scattered into the full solid angle, which substantially increases the SNR. Introducing a frequency chirp in one of the two pump beams sweeps the lattice phase velocity across the thermal velocity distribution within a single laser pulse, which allows single-shot CRBS measurements~\cite{gerakis2013single}. Alternatively, a broadband picosecond pump drives all velocities at once, and the signal is dispersed spectrally, as first predicted in~\cite{Manteghi2011SpectralNarrowing} and recently demonstrated experimentally in~\cite{Senior2026SingleshotPicosecond}.

\begin{figure}[htbp]
    \centering
    \includegraphics[width=0.6\linewidth]{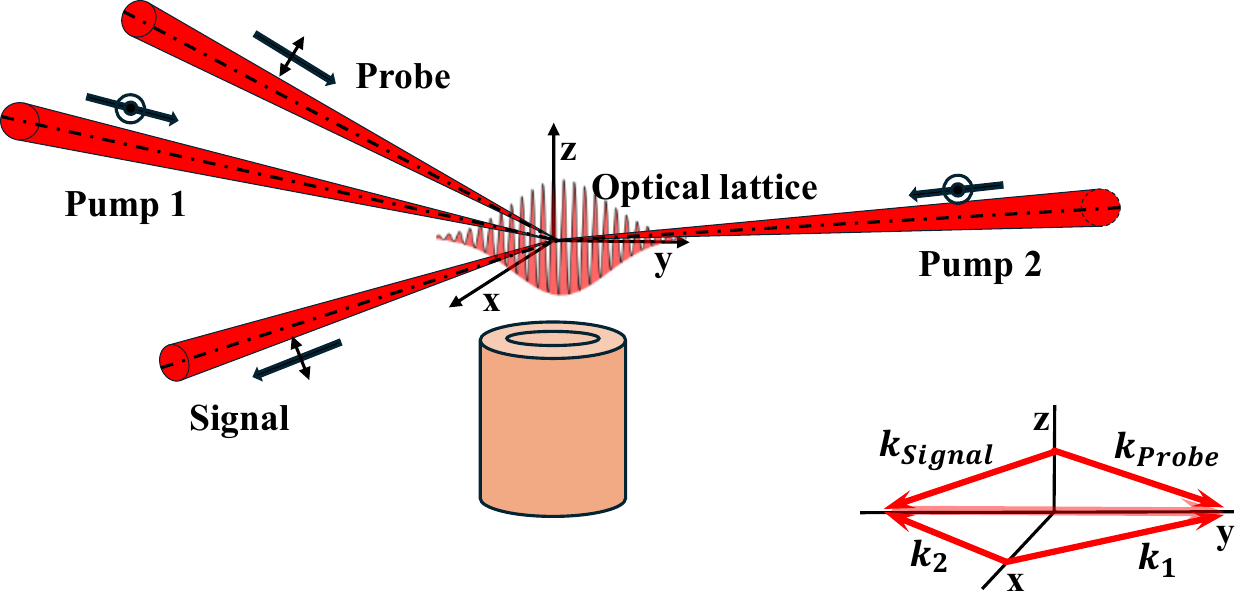}
    \caption{Beam configuration of single-shot coherent Rayleigh--Brillouin scattering. The two pump beams counter-propagate in the $xy$-plane with the same polarisation and form an optical lattice. The probe beam, with orthogonal polarisation, is incident on the lattice at the Bragg angle, ensuring that the signal beam leaves in the $yz$-plane in the phase-matched direction.}
    \label{fig:CRBScrossing}
\end{figure}

Here, the chirped-lattice approach is used. As shown in \cref{fig:CRBScrossing}, two pump beams with $\lambda_1\approx\lambda_2\equiv\lambda_\mathrm{pump}$ cross at an angle $\phi$ and form an interference pattern with period \begin{equation}
\lambda_g = \frac{\lambda_\mathrm{pump}}{2 \sin(\phi/2)}.
\end{equation}
Pump 2 has a fixed frequency $f_2$, and pump 1 is chirped such that $f_1=f_2+\Delta f(t)$, where $\Delta f$ is the frequency chirp. Polarisable scatterers experience the optical potential $U=-\frac{1}{2}\alpha E^2$ of the interference pattern, where $\alpha$ is the molecular polarisability, and the resulting dipole force $\mathbf{F} = -\nabla U$ drives them towards the high-intensity regions. The pattern, the optical lattice, moves with the phase velocity
\begin{equation}
    v_\mathrm{lat}=\lambda_g\Delta f(t),
\end{equation}
and generates a periodic density modulation, which acts as a refractive-index grating. A third beam, the \textit{probe}, is incident on this grating at an angle $\theta$ to the plane of the pumps that satisfies the Bragg condition $\lambda_\mathrm{probe}=2\lambda_g\cos\theta$, and is coherently scattered into the \textit{signal} beam, whose direction follows from the phase matching condition. The amplitude of the grating depends on the lattice velocity relative to the gas: molecules whose velocity along the grating vector is close to the lattice velocity respond most strongly, while collective (acoustic) motion adds the Brillouin response near the speed of sound. A lattice with a fixed frequency difference therefore probes the gas response at a single velocity. In single-shot CRBS, the frequency difference is chirped during the pulse so that $v_\mathrm{lat}(t)$ sweeps through the molecular velocity distribution within one pulse of about 250\,ns, and the signal is thus acquired in the time domain, allowing the use of a single photodiode.

\subsection{Laser system}
\label{ssec:laser system}
CRBS measurements are performed with a custom, portable, and frequency-agile laser system designed for single-shot CRBS, which is described in detail in Ref.~\cite{zhao2026compactdualcolourchirpedpulselaser} and based on an earlier laboratory layout~\cite{Bak2022DualAgile,Karatodorov2025HighEnergy}. In brief, a diode-pumped Nd:YVO$_4$ master laser with an intracavity electro-optic crystal, whose voltage ramp changes the cavity length and thereby chirps the laser frequency, seeds a continuous-wave slave laser at 1064\,nm. After fibre amplification, a hybrid acousto-optic/electro-optic modulator configuration cuts one pulse from an unchirped interval and one from a chirped interval of the master output. An 85\,m fibre delay line, with fine-tuning of the trigger timing, compensates the time difference between the two intervals, and both pulses are amplified in a four-stage Nd:YAG amplifier chain. The chirped pulse is split into pump 1 and the probe, and the unchirped pulse serves as pump 2, so that the frequency difference between the pumps follows the chirp. The system operates at a repetition rate of 10\,Hz and delivers pulses of approximately 250\,ns duration with energies up to 400\,mJ per arm. The chirp rate is about 15\,MHz/ns, which corresponds to a lattice acceleration of 8\,m/s per ns.

\begin{figure*}[tbp]
    \centering
    \includegraphics[width=\linewidth]{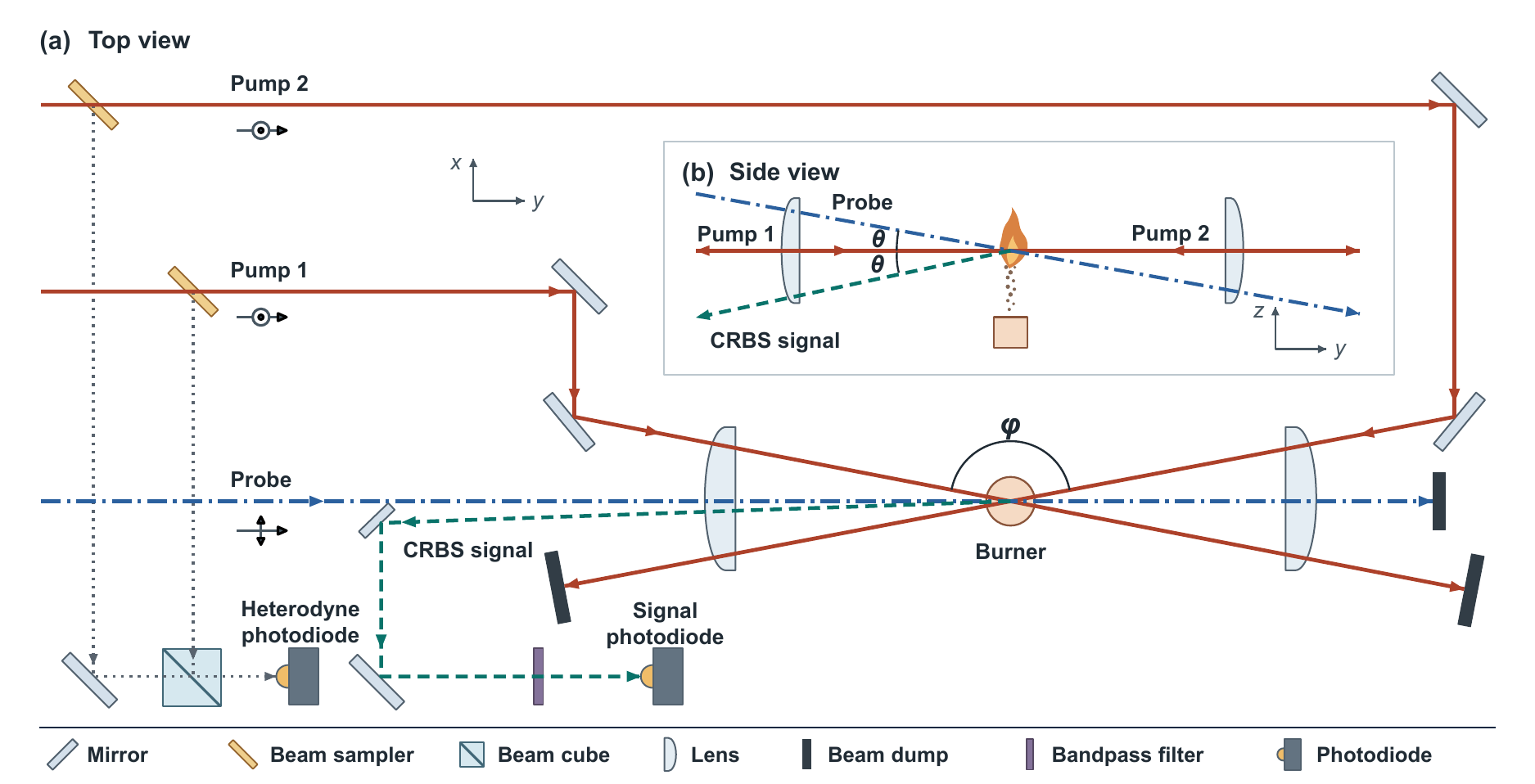}
    \caption{(a) Schematic top view of the experimental set-up. Pump 1 and pump 2 (solid red lines) counter-propagate at an angle $\phi$ in the $xy$-plane with the same polarisation, forming an optical lattice above the burner. Each pump beam is sampled (grey dotted lines) for heterodyne detection. The probe beam (blue dash-dotted line), with orthogonal polarisation, is incident on the optical lattice at the angle $\theta$ to the $xy$-plane, as shown in the side view (b). The signal beam (green dashed line) leaves at $\theta$ on the opposite side of the $xy$-plane, passes a bandpass filter, and is collected by a photodiode.
    \label{fig:scheme}}
\end{figure*}

\subsection{Optical layout and detection}
\label{ssec:optics detection}
\Cref{fig:scheme} shows the configuration of the experimental set-up. Pump 1 and pump 2 counter-propagate at an angle $\phi=176\degree$ in the $xy$-plane and are focused by a plano-convex lens (Thorlabs, LA4037-AB, focal length 400\,mm) to form the optical lattice above the burner in the horizontal plane. 

The chirp is monitored by sampling a small fraction of each pump beam and combining both through a 50:50 beam-splitter cube. Their beat signal is detected by a 5\,GHz photodiode (Thorlabs, DET08C) and recorded on a 2\,GHz oscilloscope. The probe beam, with a polarisation orthogonal to that of the pump beams to avoid interference with them, is incident at an angle $\theta=2\degree$ with respect to the $xy$-plane, corresponding to the Bragg angle, and is Bragg scattered by the optical lattice into the signal beam (inset of \cref{fig:scheme}). The signal beam leaves at $\theta=2\degree$ on the opposite side of the $xy$-plane. It is transmitted through a bandpass filter (centre wavelength 1064\,nm, FWHM $5$~nm) to suppress background light at other wavelengths. It is then focused onto a 1\,GHz photodiode (Hamamatsu, G6854-01) and recorded on the same oscilloscope as the heterodyne channel. Each shot therefore contains both the CRBS signal and its own frequency calibration.

\subsection{CRBS probe volume}
\label{ssec:probe volume}
The spatial resolution of CRBS is determined by the probe volume, where all three beams overlap and the phase-matching condition is satisfied. Its size matters here because the flame is only a few millimetres across. The probe volume is estimated from the crossing geometry, with the measured beam waist as input. The beam diameter of each beam at the common focal plane was measured with a beam profiler (Ophir, SP504S) as $\sim100$\,$\mu$m, so that the $1/e^2$ intensity radius is $w_0\approx 50$\,$\mu$m. The pump beams cross at $\phi = 176^\circ$, so that their half‐angle with the $y$-axis is $\beta = 90^\circ - \phi/2 = 2^\circ$, and the probe beam makes an angle $\theta = 2^\circ$ with the horizontal $xy$-plane. The unit direction vectors are then
\begin{align}
\mathbf{v}_1 &= (-\sin\beta,\; \cos\beta,\; 0), \label{eq:v1} \\
\mathbf{v}_2 &= (-\sin\beta,\; -\cos\beta,\; 0), \label{eq:v2} \\
\mathbf{u}   &= (0,\; \cos\theta,\; -\sin\theta). \label{eq:u}
\end{align}

\noindent At a point $\mathbf{r}$ located at a perpendicular distance $d_i(\mathbf{r})$ from the axis of a given beam $i$, the intensity of that beam is $I_i \propto \exp(-2d_i^{2}/{w_{0}^{2}})$, and the CRBS signal is proportional to the product of the two pump intensities and the probe intensity (\cref{eq:TextractionMaster}, \cref{ssec:master eq}):

\begin{equation}
S(\mathbf{r}) \propto I_{1} I_{2} I_{\mathrm{pr}}
\propto \exp\left[-\frac{2}{w_{0}^{2}}\bigl(d_{1}^{2}+d_{2}^{2}+d_{\mathrm{pr}}^{2}\bigr)\right]. \label{eq:S}
\end{equation}

\noindent On the probe axis, $\mathbf{P}=l\,\mathbf{u}$ with $l=0$ in the focal point, $d_\mathrm{pr}=0$ and 
\begin{equation}
d_{1}^{2}=d_{2}^{2}=|\mathbf{P}-(\mathbf{P}\!\cdot\!\mathbf{v}_{i})\mathbf{v}_{i}|^{2}=l^{2}\sin^{2}\beta\,(1+\cos^{2}\theta), \label{eq:d2}
\end{equation}
so that
\begin{equation}
S(l) \propto \exp\left[-\frac{4\,l^{2}\sin^{2}\beta\,(1+\cos^{2}\!\theta)}{w_{0}^{2}}\right]. \label{eq:Sprofile}
\end{equation}
Defining the probe length $L$ as the full width at which $S(l)$ drops to $1/e^2$ of its maximum gives
\begin{equation} \label{eq:L}
L = \frac{2\,w_{0}}{\sin\beta\,\sqrt{2\,(1+\cos^{2}\!\theta)}}  \approx1.4\, \mathrm{mm}.
\end{equation}
The probe volume is therefore a narrow, prolate ellipsoid (cigar-like) with a length of $\sim1.4$\,mm and a diameter of $\sim 100$\,$\mu$m. This length is comparable to the size of the flame under study (about 5\,mm), which is relevant when the flame is close to the probe volume.

\section{Data acquisition and processing}
\label{sec:data}

\subsection{Measurement campaign}
\label{ssec:campaign}
Thermometry based on the CRBS signal strength needs a calibration. Therefore, reference spectra were recorded in still room air, with no flow through the burner, at atmospheric pressure and room temperature ($T_\mathrm{ref}=297$\,K, read from the laboratory thermostat). A reference was recorded on each measurement day, numbered 1 to 4 in chronological order. 

\begin{figure}[htbp]
    \centering
    \begin{subfigure}[b]{0.25\linewidth}
        \centering
        \includegraphics[width=\linewidth]{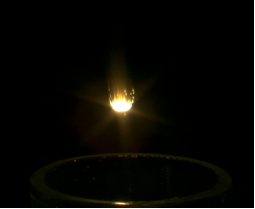}
        \caption{Flame I}
        \label{fig:FlameI}
    \end{subfigure}
    \hspace{0.01\linewidth}
    \begin{subfigure}[b]{0.25\linewidth}
        \centering
        \includegraphics[width=\linewidth]{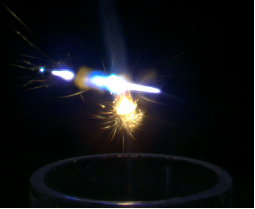}
        \caption{Flame II}
        \label{fig:FlameII}
    \end{subfigure}\\[4pt]
    \begin{subfigure}[b]{0.49\linewidth}
        \centering
        \includegraphics[width=\linewidth]{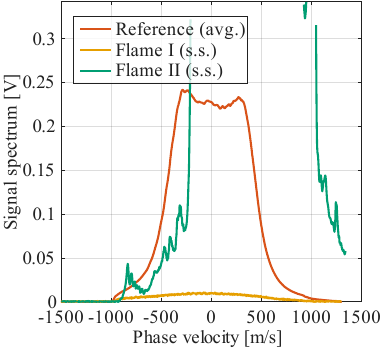}
        \caption{}
        \label{fig:CRBSflameIflameII}
    \end{subfigure}
    \caption{Typical observations during single-shot CRBS measurements on the iron powder flame. (a) Undisturbed flame; (b) intense emission and scattering during laser light--particle interaction (both 10\,ms exposure). (c) CRBS spectra against lattice velocity: an averaged room-air reference (red), a single-shot flame spectrum (yellow) corresponding to (a), and a saturated spectrum (green) corresponding to (b).}
    \label{fig:allCRBSflameIflameII}
\end{figure}

During flame operation, the data acquisition ran continuously at 10 Hz. For safety, the pump beams were blocked during manual ignition and unblocked once the flame was established. Laser light--particle interaction could not always be avoided; indeed, it is an inherent phenomenon in this particular application. Particles passing through the focused beams are inevitably irradiated. The laser light then ablates iron particles, which produces intense light emission and scattering and often extinguishes the flame, after which the ignition procedure was repeated. The flame was monitored by an area scan camera (Basler aca3088-57u) synchronised with the laser, which shows these events clearly. An example is shown in \cref{fig:FlameII}.

As the pump beams were blocked during the ignition attempts and the periods in which the flame was out, most of the about 32\,000 recorded shots, roughly 84\,\%, are blank; there was no optical lattice for the probe beam to interact with. Of the about 5\,200 flame-on shots, two thirds were affected by laser light--particle interaction and saturate the detector. Typical flame and saturated spectra are compared with the room-air reference in \cref{fig:CRBSflameIflameII}. Of the unsaturated flame-on spectra, 59 show a lineshape of gas at elevated temperature: compared with the room-air reference, the line extends over a wider range of lattice velocities, its Brillouin structure is weaker, and its central peak is much lower. These 59 spectra form the dataset of this work.

\subsection{Spectral quality}
With a chirped lattice, each lattice velocity is probed at a different time within the pulse, so the product of the pump and probe intensities at that time multiplies the spectrum point by point (\cref{eq:TextractionMaster}). A flat-top pulse is therefore essential: any temporal structure of the pulse appears as a structure in the spectrum~\cite{gerakis2013single,Bak2022DualAgile,Karatodorov2025HighEnergy}. The pulses in this campaign, however, were not ideally flat. They were shaped while monitoring the amplified pump pulses on the heterodyne photodiode, the only detector that sampled them, and this photodiode does not respond flatly to a flat optical pulse~\cite{zhao2026compactdualcolourchirpedpulselaser}. As a result, the spectra are often slightly asymmetric about their centre. No correction for the pulse shape is applied. All lineshape models are symmetric about the fitted line centre, so in an unweighted least-squares fit a weak asymmetry appears mainly in the residuals. The fitted line follows the average of the two flanks, and part of the asymmetry is taken up by the line centre offset. The chirp rate itself can cause additional asymmetry~\cite{Suzuki2024EffectsChirp,Wu2026CoherentRayleigh}, which should also be kept in mind when assessing the quality of single-shot spectra. How the pulse shape affects each temperature method is discussed in \cref{ssec:T methods}.

\subsection{From time to lattice velocity}
\label{ssec:processing}
The conversion from time to lattice velocity follows earlier single-shot CRBS work~\cite{gerakis2013single,Gerakis2016RemotesensingGas,Karatodorov2025HighEnergy} and is only summarised here, together with the choices that affect the spectral shape. The CRBS signal is obtained by subtracting the mean background trace of the same day, recorded with the pump beams blocked, and the baseline before the pulse. The instantaneous frequency follows from the analytic signal of the heterodyne trace after digital band-pass filtering from 0.125\,GHz to the oscilloscope bandwidth (2\,GHz)~\cite{Fee1992OpticalHeterodyne}. A straight line fitted to the beat frequency on both sides of its zero crossing gives the chirp rate ($\sim$15\,MHz/ns) and the crossing time. 

The signal at time $t$ then corresponds to the lattice velocity $v=\lambda_g f(t-\tau)$, where $\tau$ is the delay between the heterodyne and signal channels. Since $\tau$ is set by the detection system, it is commonly determined from the symmetry of the averaged reference spectrum and transferred to all measurements. For the flame shots, however, the delays that optimise symmetry differ from the reference delay by up to $\pm10$\,ns, whereas single reference shots scatter by only 0.5\,ns. With the reference delay, the line centres would correspond to bulk velocities up to $\pm80$\,m/s, far above the flow velocities of the burner. A net gas velocity along the grating vector would shift the spectrum in time~\cite{Gerakis2025Velocimetry, Kumar:25}, so we determine $\tau$ for each spectrum from its own symmetry through a minimisation of the difference between $S(v)$ and $S(-v)$, thereby assuming that the gas has no net velocity along the grating vector. This holds for the room-air reference and, given the low and mainly vertical flow velocities, for the flame. The symmetry point defines $v=0$ for all methods. Only the lineshape fits contain an additional line-centre offset $v_c$ as a free parameter, but it is otherwise not used.

The reference shots are averaged in velocity, each mapped with its own chirp calibration. All fits and integrals use a fixed velocity window of $-750$ to $900$\,m/s on this axis, that is, relative to the symmetry point. This window excludes the start of the pulse, where the pulse intensity and the chirp deviate most from their ideal behaviour (visible in the grey traces of \cref{fig:TfitsBoth}).

\subsection{Position of the probe volume relative to the flame}
\label{ssec:positional}
For a better insight into the combustion dynamics of iron powder flames, it would be of considerable interest to know the gas temperature as a function of the distance to the burning particles. In the current set-up, this cannot be achieved: the position of the CRBS probe volume is well known, but with a single camera monitoring the flame (\cref{fig:FlameI,fig:FlameII}), and without the flame position being stored with the spectra, the exact relative positions cannot be determined. Moreover, a non-constant saving interval of the acquisition software introduced a frame-index offset between the spectra and the camera images. Blank and saturated spectra served as markers to recover the correspondence to within a few frames.

Here, we present estimates of the distance between the probe volume and the flame position, defined as the centroid of the luminous particle cloud in the camera image (\cref{fig:FlameI}). The centroid of the probe volume was based on separately recorded reference images of a pinhole that marked the focus of the pump beams. The flame centroid relative to the pinhole was located by morphological image processing, with a scale of 36\,$\mu$m per pixel obtained from the known width of the glass tube. The camera was mounted at a slight angle to the pump-beam axis due to limited space, and the estimate is a two-dimensional projection that ignores motion along the line of sight. As the flame moves in all three directions, these estimates (2.8--12.3\,mm, median 7.3\,mm) carry considerable uncertainty. Between consecutive frames (100\,ms apart), the estimated distance changes by only 0.2\,mm in the median and 1.0\,mm at most, so any remaining frame-index offset has little effect. Moreover, the 1.4\,mm long probe volume may contain temperature gradients when the flame is close. Since the signal scales as $T^{-4}$ (\cref{eq:TextractionMaster}), colder gas contributes more, so that each measured temperature is a signal-weighted average biased towards the colder part of the probe volume. Future work needs a three-dimensional tracking approach to fully characterise the gas temperature distribution of an iron powder flame.

\section{Temperature extraction methods}
\label{sec:Textractionmethods}

\subsection{Signal strength and lineshape}\label{ssec:master eq}
For a weak grating (assumption A4 in Appendix~\ref{app:derivation}), the diffracted probe power is proportional to the square of the modulation of the dielectric constant, $S\propto|\delta\varepsilon|^2I_\mathrm{pr}$~\cite{BOYD2020419,Gerakis2016RemotesensingGas,Pan2004CoherentRayleighBrillouin}. In a dilute gas, $\varepsilon_r=1+n\alpha/\varepsilon_0$, so that a number density grating with amplitude $\delta n$ gives $\delta \varepsilon=\alpha\delta n/\varepsilon_0$~\cite{BOYD2020419}. The density grating is driven by the optical potential of the lattice, with amplitude
\begin{equation}\label{eq:TextractionPotential}
    U_0 = \dfrac{\alpha \sqrt{I_1I_2}}{c\varepsilon_0}.
\end{equation}
For a weak drive ($U_0\ll k_BT$), the response is linear in $U_0/k_BT$, and the density amplitude can be written as:
\begin{equation}
\delta n = n_0\,\frac{U_0}{k_B T}\,G\!\left(\xi,y\right),\qquad \xi=\frac{v}{v_0},\qquad v_0=\sqrt{\frac{2k_B T}{m}},
\label{eq:dn}
\end{equation}
where $n_0$ is the number density, $m$ the molecular mass, $v_0$ the most probable speed, and $G$ the complex response of the gas to a lattice that moves with velocity $v$ relative to the gas (see Appendix~\ref{app:derivation}). $G$ depends on the dimensionless velocity $\xi$ and on the uniformity parameter
\begin{equation}
    y = \frac{p}{k_g \eta_s v_0},
    \label{eq:yparam}
\end{equation}
which compares the grating period with the mean free path. Here $p$ is the pressure, $k_g=2\pi/\lambda_g$ the grating wavenumber, and $\eta_s$ the shear viscosity. For molecular gases, $G$ also depends on the relaxation of internal energy, through the bulk viscosity $\eta_b$ and the thermal conductivity $\kappa$~\cite{Pan2004CoherentRayleighBrillouin}. Combining these relations gives:
\begin{equation}
S(v)=C\,\alpha^4\left(\frac{n_0}{T}\right)^{2}I_1I_2I_\mathrm{pr}\;\Phi(\xi,y),\qquad \Phi\equiv|G|^2,
\label{eq:TextractionMaster}
\end{equation}
where $C$ contains the interaction geometry, detection response, and physical constants. Two powers of $\alpha$ come from driving the grating, and two from reading it out. In a single shot, $I_1I_2I_\mathrm{pr}$ is the value at the time at which velocity $v$ is probed. 

\Cref{eq:TextractionMaster} separates the signal into a \textit{strength}, $\alpha^4(n_0/T)^2$, and a \textit{shape}, $\Phi(\xi,y)$. Two exact properties of $\Phi$ are central to this work (see Appendix~\ref{app:derivation}): 
\begin{align}
\Phi(0,y)&=1 \quad\text{for every } y, \label{eq:Phi xi 0}\\
\Phi(\xi,0)&=\bigl|1+\xi Z(\xi)\bigr|^2, \label{eq:Phi y 0}
\end{align}
where $Z$ is the plasma dispersion function~\cite{Fried1961PlasmaDispersion,Pan2004CoherentRayleighBrillouin}. \Cref{eq:Phi xi 0} states that a lattice at rest relative to the gas produces the static Boltzmann redistribution $\delta n/n_0=U_0/k_BT$, independent of collisions. The same static amplitude follows from the continuum theory of electrostriction~\cite{BOYD2020419} (Appendix~\ref{app:derivation}). \Cref{eq:Phi y 0} is the collisionless limit. The lineshape between these two limiting cases follows from a kinetic model. Here we use the S6 model of Pan et al.~\cite{Pan2004CoherentRayleighBrillouin}. It solves the linearised Wang Chang--Uhlenbeck equation with a six-moment collision model and is the coherent counterpart of the Tenti S6 model for spontaneous Rayleigh--Brillouin scattering~\cite{Pan2004CoherentRayleighBrillouin,Tenti1974KineticModel}.

At constant pressure, $n_0=p/k_BT$, so the signal scales as $T^{-4}$ at a fixed lineshape. At fixed temperature, \cref{eq:TextractionMaster} contains the quadratic dependence on density that Gerakis et al.\ verified for the spectrally integrated signal in several gases~\cite{Gerakis2016RemotesensingGas}. The additional factor $T^{-2}$ comes from the driven amplitude $U_0/k_BT$, and it is what makes the CRBS signal strength a sensitive thermometer at constant pressure.

\subsection{Four temperature extraction methods}\label{ssec:T methods}
The four methods rely on different parts of \cref{eq:TextractionMaster} and on different assumptions, so that comparing them tests the signal model and shows which assumptions limit the accuracy. Two methods use the signal \textit{strength} of a flame spectrum and compare it to the calibration reference spectrum recorded in still room air at known pressure, temperature, and composition (\cref{ssec:campaign}). They assume equal pressure, and equal pulse energies, pulse shapes, and beam overlap in both measurements. Equal beam overlap, however, is not guaranteed in a flame, because refractive-index gradients can steer the beams and particles can attenuate them. The other two methods only use the \textit{shape} of the spectrum. For them, $C\alpha^4(n_0/T)^2I_1I_2I_\mathrm{pr}$ reduces to a single fitted amplitude $A$.

\subsubsection{M1: signal at zero velocity}
\label{sssec:M1}
At zero lattice velocity, \cref{eq:Phi xi 0} removes the lineshape from \cref{eq:TextractionMaster}, and with $n_0=p/k_BT$:
\begin{equation}
    S(0) \propto \dfrac{\alpha^4p^2}{T^4}I_1I_2I_\mathrm{pr}.
\end{equation}
Using a reference calibration spectrum at constant pressure gives
\begin{equation}
    T = T_\mathrm{ref}\dfrac{\alpha}{\alpha_\mathrm{ref}}\left[\dfrac{S_\mathrm{ref}(0)}{S(0)}\right]^{1/4}.
\end{equation}
This method needs no lineshape model, no transport coefficients, no molecular mass, and no chirp calibration beyond locating $v=0$. Since the signal at zero velocity is measured at one instant in the pulse, M1 depends only on the pump and probe intensities at that instant, relative to the reference, and not on the pulse shape elsewhere. In principle, it does not even require a spectrally resolved measurement: a lattice at rest, formed by two pump beams of equal frequency from a seeded, narrowband laser, would suffice. Such a lattice is at rest relative to the gas only if the bulk velocity along the grating vector is small compared with $v_0$, since a bulk flow shifts the spectrum~\cite{Gerakis2025Velocimetry,Kumar:25}. Since the lineshape is symmetric in $\xi$, a bulk velocity $v_b$ changes $S(0)$ to the second order in $v_b/v_0$, so that the error is small for $v_b\ll v_0$, but grows when the bulk velocity approaches the thermal velocity. 

The temperature scales with the polarisability, and a relative error in $S(0)$ enters that of the temperature with a factor of 1/4. In this work, $S(0)$ is obtained from a second-order polynomial fitted to the spectrum within $\pm$200\,m/s and evaluated at $v=0$. $S_\mathrm{ref}(0)$ is obtained in the same way from the averaged reference. Because the polynomial contains a linear term, a local slope of the spectrum at $v=0$, as caused by a weak asymmetry, does not affect $S(0)$. The order and range of this polynomial are not critical, and any smooth function that describes the data around $v=0$ would serve. A second-order polynomial over $\pm$200\,m/s describes the spectra well in this range, and the range is wide enough to average out most of the noise.

\subsubsection{M2: velocity integral}\label{sssec:M2}
Integrating \cref{eq:TextractionMaster} over a velocity window $W$ gives $\int_W S\,\mathrm{d}v = S(0) K_W(T)$, with the line area $K_W(T)=\int_W\Phi\,\mathrm{d}v$. The temperature follows from
\begin{equation}
\dfrac{\int_W S_\mathrm{ref}\,\mathrm{d} v}{\int_W S\,\mathrm{d} v}=\left(\dfrac{\alpha_\mathrm{ref}}{\alpha}\right)^{4}\left(\dfrac{T}{T_\mathrm{ref}}\right)^{4}\dfrac{K_W(T_\mathrm{ref})}{K_W(T)},
\label{eq:TextractionAreaRatio}
\end{equation}
which is solved numerically for $T$. The integral uses the same fixed window as the fits, without any velocity shift, and $K_W$ is computed with the S6 model with the same settings as M4. For the full line, $K=v_0\int\Phi\,\mathrm{d}\xi$ grows as $\sqrt{T}$, apart from the dependence on $y$. In the window used here, the effective exponent of $T$ is close to 4. Using the area of the full line instead would overestimate the temperature, because a hotter line loses a large fraction of its wings outside the window. Compared with M1, this method uses the whole measured window, which reduces the noise, but it needs the lineshape model, the mass, and a flat pulse envelope over the line. For the flames, $K_W$ uses the $\eta_b/\eta_s$ of the M4 fit of the same spectrum, and for the reference, the value fitted to the averaged reference at $T_\mathrm{ref}$.

\subsubsection{M3: collisionless lineshape}
For $y\to0$, the lineshape is \cref{eq:Phi y 0}, and the CRBS spectrum is fitted with
\begin{equation}
S(v)=A\,\left|1+\xi Z(\xi)\right|^2,\qquad \xi=\frac{v-v_c}{\sqrt{2k_B T/m}},
\label{eq:TextractionCollisionlessFit}
\end{equation}
with $A$, $T$, and the line-centre offset $v_c$ free. Since $Z$ is a fixed mathematical function, this fit needs only the molecular mass, and the temperature is proportional to the assumed $m$. For this study, however, this method is expected to be least accurate, yet it is included because it is the exact limit for $y\to0$ and needs no transport coefficients. In practice, the flames have $y>0$, collisions flatten the line centre, and the collisionless fit overestimates the temperature. The size of this bias depends on $y$ and on the part of the line inside the velocity window. Although a higher temperature lowers $y$, a fixed window then captures a smaller fraction of the broader line, so the accuracy does not simply improve with temperature. Recording the full line, with broadband pumps~\cite{Senior2026SingleshotPicosecond} or a larger chirp range, would reduce the bias at the present flame temperatures, but the collisional part vanishes only for $y\to0$. At atmospheric pressure, this requires very high gas temperatures: in air, $y=0.1$ is reached only at about 2600\,K, whereas reducing the pressure lowers $y$ without broadening the line. The collisionless approach is therefore best suited to measurements at reduced pressure. The expected bias is evaluated in \cref{ssec:fulldataset}. 

\subsubsection{M4: kinetic lineshape}\label{sssec:kinetic}
The spectrum is fitted with $S(v)=A\,\Phi_\mathrm{S6}(\xi,y)$, with $\xi=(v-v_c)/v_0$ as in \cref{eq:TextractionCollisionlessFit}, and with $A$, $T$, $v_c$ and the ratio of bulk to shear viscosity $\eta_b/\eta_s$ free. The shear viscosity and thermal conductivity are those of N$_2$ from Sutherland's law~\cite{Gu2014SystematicStudy}, with the corresponding Eucken factor. The bulk viscosity is the least characterised transport coefficient~\cite{Gu2013TemperaturedependentBulk,Li2025GasViscosity,Pan2004CoherentRayleighBrillouin}, and its ratio to the shear viscosity is fitted within $0.01$--$10$. This is the only method that describes the Brillouin structure, and it contains M3 as its limit $y\to0$.

\subsubsection{Implementation}
The gas around the CRBS probe volume is a mixture of \ch{N2}, \ch{O2} and possibly combustion products, in proportions that are not precisely known. It is treated as a single-component gas with an effective mass and polarisability: for the flame gas, those of the 50/50 \ch{O2}/\ch{N2} shroud mixture ($m=30.0$\,u, $\alpha=1.661$\,\AA$^3$), and for the reference, those of dry air ($m=28.96$\,u, $\alpha=1.707$\,\AA$^3$). Oxygen consumption by the burning particles is not accounted for. The four methods and the input each requires are summarised in \cref{tab:methods}. All fits are unweighted least-squares fits on the measured, non-uniformly spaced velocity samples within the fixed window.

\begin{table}[htbp]
\small
\centering
\caption{Temperature extraction methods and the input each requires.}
\label{tab:methods}
\begin{tabular}{@{}lcccc@{}}
\toprule
                           & M1: $S(0)$ & M2: $\int S\,\mathrm{d} v$ & M3: collisionless fit & M4: S6 fit          \\ \midrule
Absolute signal      & yes        & yes                        & no                    & no                  \\
Lineshape model            & no         & yes ($K_W$)                & limit only            & yes                 \\
Temperature via & $T^{-4}$   & $T^{-4}K_W(T)$             & $v_0$           & $v_0$ and $y$ \\
Transport coefficients     & no         & yes                        & no                    & yes                 \\
Composition enters as      & $\alpha$   & $\alpha$, $m$, transport            & $m$                   & $m$, transport      \\
Pulse shape over the line  & no         & yes                        & yes                   & yes                 \\ \bottomrule
\end{tabular}
\end{table}

\section{Results and discussion}\label{sec:results}
\subsection{Single-shot spectra and fit quality}\label{ssec:ss spectra}
\begin{figure*}[tbp]    
    \centering    
    \begin{subfigure}{0.49\textwidth}
        \centering
        \includegraphics[width=\linewidth]{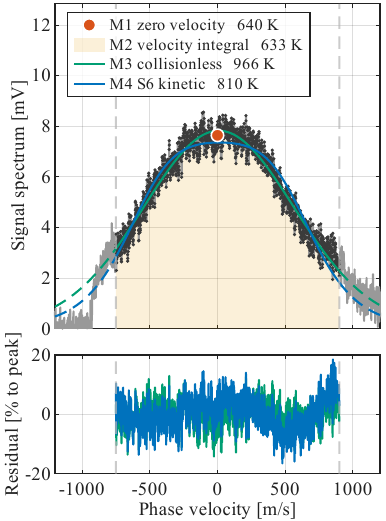}
        \caption{Spectrum A}
        \label{fig:Tfit888}
    \end{subfigure}\hfill
    \begin{subfigure}{0.49\textwidth}
        \centering
        \includegraphics[width=\linewidth]{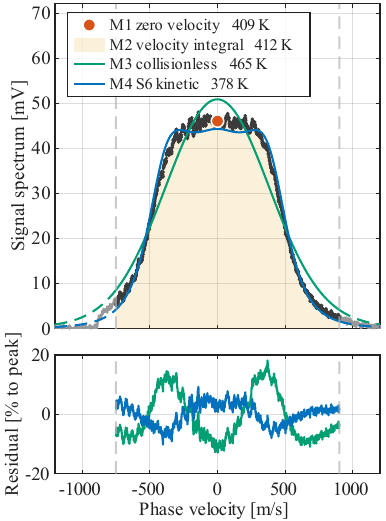}
        \caption{Spectrum B}
        \label{fig:Tfit643}
    \end{subfigure}    
    
    \captionsetup{width=\textwidth}
    \caption{Two single-shot CRBS spectra with the results of the four temperature methods. (a)~Spectrum A shows the spectrum that gives the highest temperature in the dataset with $y = 0.36$. (b)~Spectrum B, a spectrum of colder gas with $y = 0.88$. The grey line shows the full trace, and only the dotted data between the dashed vertical lines are used in the fit and integration window. The large red dot at $v=0$ is the signal $S(0)$ used by M1. The shaded area depicts the window integral used by M2. The green and blue curves are the M3 and M4 fits, dashed outside the window. The lower panels show the fit residuals in percent of the peak signal.}
    \label{fig:TfitsBoth}
\end{figure*}

The two single-shot spectra in \cref{fig:TfitsBoth} show example flame spectra and the corresponding temperatures from all four methods. Spectrum A is the spectrum in the dataset with the highest temperature, with an M4 temperature of 810\,K, corresponding to $y=0.36$. Spectrum B is one of the spectra with the lowest temperature and gives an M4 temperature of 378\,K and $y=0.88$. The peak signal of spectrum A is six times lower than that of spectrum B, so noise dominates the residuals of both fits. The flat top in the signal of spectrum B is a remnant of the merged Brillouin peaks (indicating a higher $y$) and is reproduced by the S6 model with an rms residual of about 4\,\% of the peak. 

The coherent S6 fits describe the spectra well, but not perfectly. The roughly symmetric residual pattern shows that the modelled Brillouin peaks are too prominent and the central Rayleigh peak is underestimated. A similar response by the kinetic model was reported by Pan et al.\ for some of their measurements~\cite{Pan2004CoherentRayleighBrillouin}. The collisionless fit shows the opposite pattern, where the fit is too peaked in the centre and too wide in the wings. The measured lines thus lie between the collisionless and the S6 shape, and closer to the latter. 

Several explanations for the misfits can be examined with the data. The composition of the flame gas cannot be the sole contributor since the room-air references, whose composition and temperature are known, exhibit the same pattern. For that reason, a temperature non-uniformity within the probe volume is unlikely to be the main cause either, although a superposition of lines of different widths could, of course, produce a similar pattern. The misfit must therefore originate in the measurement itself or in the model. Both possibilities are discussed below. 

On the side of the measurement, an unresolved instrument function would smear the Brillouin structure and make the line appear closer to collisionless. To reproduce the reference spectra with literature transport coefficients, a Gaussian instrument function with a standard deviation of 60--70\,m/s would be needed, for which we find no physical source. On the model side, the S6 lineshape at a given $y$ depends on the shear viscosity through $y$ itself. If the shear viscosity of the model is increased by a factor chosen such that each reference is fitted at its known temperature, this factor becomes 1.1--1.3 and the reference residuals decrease from 2.9--3.9\,\% to 1.1--2.7\,\%. Applied to the flames, the same factors remove most of the central misfit and reduce the median residual to within 3\,\%. The temperature increases by 1.5\,\% in the median, at most by 22\,K. Such an effective viscosity has no physical justification, since the shear viscosities of \ch{N2} and air are known~\cite{White2006ViscousFluid}. Nor is there a reason why a correction found in air at room temperature would apply as well to the flame gas at 400--800\,K. We therefore do not apply this correction, and regard the resulting temperature change of about 1.5\,\% as a systematic uncertainty of the lineshape model.

The upper bound of 10 on $\eta_b/\eta_s$ (\cref{sssec:kinetic}) deserves a comment, because its value is uncertain. The bulk viscosity of \ch{N2} and \ch{O2} from acoustic measurements corresponds to ratios of 0.73 and 0.4, and that of \ch{CO2} to a ratio of about 1000~\cite{Pan2004CoherentRayleighBrillouin}. Pan et al.\ note that these values were measured at megahertz frequencies and may not apply at the gigahertz frequencies of CRBS~\cite{Pan2004CoherentRayleighBrillouin}. It is also noted that for ratios of order 1000, the modelled Brillouin peak positions become insensitive to the internal heat capacity~\cite{Pan2003CoherentRayleighBrillouin}. The internal degrees of freedom are then effectively frozen on the time scale of the lattice. To stay clear of this regime, we limited the ratio to 10. In the references, the fitted ratio is 1.9--2.4, but in all flame spectra it reaches the upper bound. Above a ratio of about 5, the S6 lineshape saturates, and between 10 and 1000 the flame temperatures change less than 0.25\,\%. The ratio is thus not determined by the flame spectra, but neither does it bias $T_\mathrm{M4}$.

\subsection{Temperatures of the full dataset}\label{ssec:fulldataset} 
\begin{figure}[htbp]
    \centering
    \includegraphics[width=0.5\linewidth]{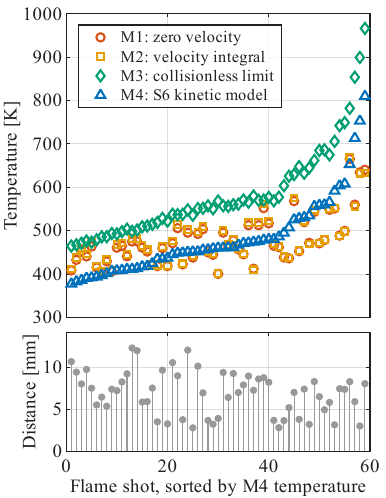}
    \caption{Temperatures of all 59 single-shot spectra from the four methods, sorted by the M4 temperature (top), and the corresponding estimated distance between the flame centroid and the centre of the probe volume (bottom).}
    \label{fig:Tresultsmethods}
\end{figure}

\Cref{fig:Tresultsmethods} shows the temperatures of all 59 spectra from the four methods, together with the estimated distance between the flame centroid and the probe volume. The spectra are sorted by M4 temperature, the only method that uses the full lineshape and needs no reference measurement. The M4 temperature ranges from 378 to 810\,K. M1 and M2 give 400--669\,K, and M3 gives 465--966\,K. The temperature varies strongly from shot to shot, although the operating conditions were nominally identical. Consecutive recorded shots differ by up to 330\,K in M4 and 163\,K in M2. These changes do not depend on the reference calibration, since consecutive shots share the same reference, and the methods agree on them if the change is large: all eight consecutive pairs with a change of more than 50\,K in $T_\mathrm{M2}$ show a change of the same sign in $T_\mathrm{M4}$. The estimated distance between the flame centroid and the probe volume, as stated in \cref{ssec:positional}, ranges from 2.8 to 12.3\,mm and does not explain this variability. Its correlation with temperature is weak for all methods, which is consistent with the large uncertainty of the two-dimensional estimate of a flame that moves in three dimensions. The temperature variability is therefore not explained by a single spatial variable and should instead be interpreted as a consequence of the inherent unsteadiness and inhomogeneity of the iron powder flame. Every laser pulse samples a different instantaneous gas environment, and a time-averaged measurement would suppress this variability and yield a temperature that is unrepresentative of any individual flame state. This shows that single-shot acquisition is essential for dynamic gas temperature measurements in iron powder flames, and that single-shot CRBS captures these fluctuations.

\begin{figure*}[tbp]    
    \centering    
    \begin{subfigure}{0.49\textwidth}
        \centering
        \includegraphics[width=\linewidth]{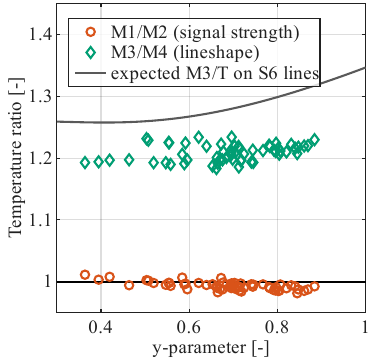}
        \caption{}
        \label{fig:ratios_vs_y}
    \end{subfigure}\hfill
    \begin{subfigure}{0.49\textwidth}
        \centering
        \includegraphics[width=\linewidth]{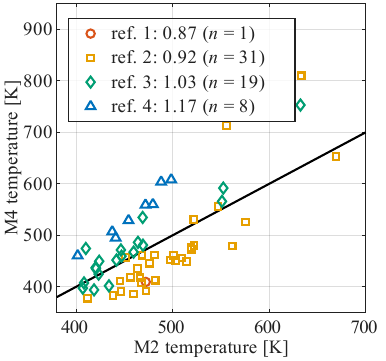}
        \caption{}
        \label{fig:M4_vs_M2}
    \end{subfigure}    
    \captionsetup{width=\textwidth}
    \caption{Agreement within (a) and between (b) the pairs of methods. (a) Temperature ratios within the pairs against $y$ from the M4 fit. The grey line shows the expected ratio for noise-free S6 lines fitted with the collisionless lineshape in the same fixed $-750$ to $900$\,m/s window. (b) M4 temperatures are plotted against M2 temperatures for all flame spectra, grouped by the reference recording used for M2. The line marks equality. For each reference, the grouped median M4/M2 and the number of spectra $n$ are indicated.}
    \label{fig:Tpairs}
\end{figure*}

Moreover, two features stand out in the temperature results. M1 and M2 nearly coincide for every spectrum, and M3 lies above M4 by an almost constant factor. Between the M1/M2 pair and the M3/M4 pair, the agreement varies from spectrum to spectrum. \Cref{fig:ratios_vs_y} shows the temperature ratios within the two pairs.

The signal-strength methods agree within 2\,\% for every spectrum, with $T_\mathrm{M1}/T_\mathrm{M2}=0.99$ in the median. This agreement is not trivial. M1 uses only the signal at zero velocity, while M2 integrates the whole spectrum and needs the modelled line area. It shows that the shape of the flame spectra relative to the reference is consistent with \cref{eq:TextractionMaster}. Both methods, however, normalise the flame signal by the same reference signal. A systematic error in this absolute calibration would shift M1 and M2 equally, so their agreement cannot reveal it.

The lineshape methods themselves differ by an almost constant factor of 1.21, with all spectra between 1.18 and 1.23. This is the expected bias inherent to using M3 outside the collisionless limit. A correction needs the S6 or equivalent model, in which case the corrected M3 would simply become M4. Over the $y$ range of the flames and with the current velocity window, the expected ratio is 1.26--1.32 (grey line in \cref{fig:ratios_vs_y}). The measured ratios follow its weak dependence on $y$, but lie about 6\,\% lower. This is consistent with the residuals of \cref{ssec:ss spectra}, where the measured lines were found to be slightly closer to the collisionless shape than the S6 model predicts.

Between the pairs, the ratio $T_\mathrm{M4}/T_\mathrm{M2}$ is also 0.99 in the median, but ranges from 0.83 to 1.28. \Cref{fig:M4_vs_M2} shows that this spread is ordered largely by the reference recordings used for M1 and M2. Within each reference group, M4 and M2 are strongly correlated, with Pearson correlation coefficients of 0.85, 0.95, and 0.98 for references 2 to 4. The median ratio, however, is 0.92, 1.03, and 1.17 for these groups. The methods thus follow the same changes from spectrum to spectrum, while the absolute scale of M2 depends on the reference. The references themselves are stable, with a shot-to-shot standard deviation of at most 2.6\,\% in $S(0)$ within each recording. Their absolute levels, however, range from 114 to 224\,mV.

The cause of this calibration transfer cannot be identified from the present data. The optical alignment was not changed during the campaign, and all references were recorded in still room air. Possible causes are differences in pulse energy and temporal pulse shape between the reference and the flame recordings, which are not separately monitored for every shot, and attenuation or steering of the beams by the particle-laden flame, which would affect only the flame spectra. This transfer is the dominant systematic uncertainty of M1 and M2 in this dataset. It is a property of the measurement, not of the method.

\section{Conclusion}\label{sec:conclusion}
Single-shot coherent Rayleigh--Brillouin scattering was used to characterise a small laboratory-scale iron powder flame at atmospheric pressure. Of about 5\,200 flame-on shots, 59 yielded spectra of hot gas without laser light--particle interaction; most of the others were saturated by this interaction. From one expression for the CRBS signal, \cref{eq:TextractionMaster}, four temperature methods were derived and applied to each spectrum: the signal at zero lattice velocity (M1) and the velocity-integrated signal (M2), both referenced to room air at constant pressure, and fits of the coherent collisionless (M3) and kinetic S6 (M4) lineshape. 

The four methods form two consistent pairs. M1 and M2 agree within 2\,\% for every shot, which shows that the shape of the flame spectra relative to the reference is consistent with the signal model. M3 exceeds M4 by a roughly constant 20\,\%, somewhat below the bias expected for a collisionless fit at $y=$0.4--0.9 in the present window. The collisionless fit is transport-free, but becomes accurate only at much lower $y$, which is best reached at reduced pressure and unrealistic for combustion applications. It is included here as the limiting case of the kinetic model.

Between the pairs, the calibration of the signal strength against room-air references is the dominant systematic uncertainty of the signal-strength methods in this dataset. Since M1 and M2 rely on the absolute signal, they are also directly affected by changes in beam overlap, such as beam steering by refractive-index gradients, to which the lineshape methods are insensitive.

M1 nevertheless has distinct advantages. Since $\Phi(0,y)=1$ for any conserving collision model, it requires no lineshape model and no transport coefficients, and the composition enters only through the polarisability, whereas M4 requires the mass and transport coefficients of the mixture. In principle, M1 does not even require a spectrally resolved measurement, only a lattice at rest relative to the gas,  which a seeded, narrowband laser can provide when the bulk flow along the grating vector is small compared with the thermal velocity. M1 is therefore attractive for gases of known composition, provided the beam overlap and pulse energies are stable between reference and measurement. To our knowledge, the signal at zero lattice velocity has not been used for gas thermometry before.

The gas temperature near the flame ranged from about 400 to 800\,K and changed by up to several hundred kelvin between consecutive shots. Time-averaged diagnostics are therefore unsuitable for this flame, and single-shot acquisition is essential.

Future work should (i) ensure flat-top pulses on a detector with a verified flat response, and monitor the energies and temporal shapes of all beams individually for every shot, (ii) record reference spectra interleaved with the flame measurements, (iii) track the centroid of the burning particle cloud relative to the probe volume centre in three dimensions, and (iv) extend the measurements to larger flames with lower local particle concentrations, where measurements within the particle cloud may become possible.

\appendix
\section{Derivation of the signal expression}
\label{app:derivation}
This appendix derives \cref{eq:TextractionMaster} and the properties (\ref{eq:Phi xi 0}) and (\ref{eq:Phi y 0}). The assumptions are: (A1) a weak drive of the gas by the pump beams, such that the gas density responds linearly to the optical potential, $U_0/k_B T\ll1$; (A2) a quasi-steady response, i.e.\ the gas responds within $1/(k_gv_0)\approx0.2$\,ns, during which the chirp moves the lattice by about 1.6\,m/s, much less than $v_0$; (A3) collisions that conserve particle number, momentum, and energy, and a single polarisability per molecule; and (A4) a weak grating, so that only a small fraction of the probe is diffracted and the diffracted power is proportional to $|\delta\varepsilon|^2$.

Following Pan~\cite{Pan2004CoherentRayleighBrillouin}, $z$ in this appendix denotes the coordinate along the grating vector, i.e.\ the $y$-axis of \cref{fig:CRBScrossing}.

\paragraph{Optical potential.} Two pump fields with the same polarisation act on a molecule through $U=-\tfrac12\alpha E^2$. After averaging over the optical period, the cross term forms a lattice moving with $v=\omega/k_g$, where $k_g=|\mathbf k_1-\mathbf k_2|$ and $\omega=\omega_1-\omega_2$:
\begin{equation}
U(z,t)=-U_0\cos(k_gz-\omega t),\qquad U_0=\tfrac12\alpha E_{10}E_{20}=\frac{\alpha\sqrt{I_1I_2}}{c\,\varepsilon_0}.
\end{equation}

\paragraph{Kinetic equation.} Write the velocity distribution as $f=n_0f_0(u)[1+h(z,u,t)]$, with $f_0$ the Maxwellian and $u$ the molecular velocity along $z$. Linearised in the drive (A1), the Boltzmann equation reads~\cite{Pan2004CoherentRayleighBrillouin}
\begin{equation}
\frac{\partial h}{\partial t}+u\frac{\partial h}{\partial z}+\frac{2u}{v_0^2}\,\frac{1}{m}\frac{\partial U}{\partial z}=\mathcal C[h],
\end{equation}
with $\mathcal C$ the linearised collision operator. For $h=\mathrm{Re}\{\hat h\,e^{i(k_gz-\omega t)}\}$ and $v_0^2=2k_B T/m$:
\begin{equation}
ik_g(u-v)\,\hat h-ik_g\,u\,\frac{U_0}{k_B T}=\mathcal C[\hat h].
\label{eq:kin2}
\end{equation}
The drive enters only through $U_0/k_B T$, so $\hat h\propto U_0/k_B T$, and the density amplitude is $\delta\hat n=\int n_0f_0\hat h\,\mathrm{d} u=n_0(U_0/k_B T)\,G$. In the variable $s=u/v_0$, \cref{eq:kin2} contains only $\xi=v/v_0$ and the collision rate in units of $k_gv_0$, i.e.\ $y$; for molecules, $G$ also depends on the internal relaxation parameters of the S6 kinetic model.

\paragraph{Static lattice.} At $\xi=0$, the constant $\hat h=U_0/k_B T$ solves \cref{eq:kin2}: the left-hand side vanishes, and $\mathcal C$ of a constant is zero by (A3), since a constant $h$ is a Maxwellian of a different density. This is the linearised Boltzmann distribution $n=n_0\exp(U_0\cos k_gz/k_B T)$, hence $G(0,y)=1$ for every $y$ and every conserving collision model. The same static amplitude follows from the continuum theory of electrostriction~\cite{BOYD2020419}: $\delta n/n_0=\tfrac12\varepsilon_0C_T\gamma_eE_{10}E_{20}$, which for an ideal gas, with compressibility $C_T=1/p$ and electrostrictive constant $\gamma_e=n_0\alpha/\varepsilon_0$, is exactly $U_0/k_B T$.

\paragraph{Collisionless limit.} With $\mathcal C=0$, $\hat h=(U_0/k_B T)\,u/(u-v)$, and the velocity integral gives $G(\xi,0)=1+\xi Z(\xi)$, with $Z$ the plasma dispersion function.

\paragraph{Signal.} With $\delta\varepsilon=\alpha\,\delta n/\varepsilon_0$ and $S\propto|\delta\varepsilon|^2I_\mathrm{pr}$ (A4)~\cite{BOYD2020419,Gerakis2016RemotesensingGas,Pan2004CoherentRayleighBrillouin},

\begin{equation}
S\propto\frac{\alpha^2}{\varepsilon_0^2}\,n_0^2\frac{U_0^2}{k_B^2T^2}|G|^2I_\mathrm{pr}\propto\alpha^4\frac{n_0^2}{T^2}\,I_1I_2I_\mathrm{pr}\,|G|^2,
\end{equation}
which is \cref{eq:TextractionMaster}. In Pan's formulation~\cite{Pan2004CoherentRayleighBrillouin}, which studies the lineshape of a single gas, the read-out factor $\alpha^2$ is a constant and only $|\delta n|^2$ appears.

\begin{backmatter}
\bmsection{Funding}
Content in the funding section will be generated entirely from details submitted to Prism. Any text added to this section will be replaced during production and will display official funder names along with any grant numbers provided. 

\bmsection{Acknowledgment}
The authors thank Viktor Kornilov for his help with the preparation of the burner and Mikhail Shneider for discussions on the shear and bulk viscosity in the coherent S6 model. Y.Z. and A.G. are supported by the Luxembourg National Research Fund 15480342 (FRAGOLA).

\bmsection{Disclosures}
The authors declare no conflicts of interest.

\bmsection{AI disclosure}
The authors used Claude (Opus 5.5) to assist in optimising MATLAB scripts used for data processing, to check the derivation in Appendix~A, for grammar refinement, and for organisational suggestions within the manuscript. All processing code was verified against independent implementations. 

\bmsection{Data availability}
Data underlying the results presented in this paper are not publicly available at this time but may be obtained from the authors upon reasonable request.

\end{backmatter}

\bibliography{LIST-TUE-refs}

\end{document}